\documentclass[useAMS,usenatbib]{mn2e}
\usepackage{graphicx}
\usepackage{txfonts}
\usepackage{wrapfig}
\usepackage{longtable}
\usepackage[usenames]{color}

\def\h2{\rm{H_2}}
\def\fh2{f_{\rm{H_2}}}
\def\dfHI{\Delta\rm{f_{HI}}}
\def\mHI{M_{\rm{HI}}}
\def\mh2{M_{\rm{H2}}}
\def\SHI{\Sigma_{\rm{HI}}}
\def\Sh2{\Sigma_{\h2}}
\def\sgas{\Sigma_{\rm{gas}}}

\def\ms{M_{\odot}}

\def\HIs{M_{\rm{HI}}/M_*}

\def\mspc{M_{\odot}\rm{pc^{-2}}}

\def\rs{r_{\rm s}}
\def\rc{r_{\rm c}}

\def\aj{AJ}
\def\apj{ApJ}
\def\apjl{ApJ}
\def\apjs{ApJS}
\def\aap{A\&A}
\def\mnras{MNRAS}

\begin{document}
\title[Bluedisk II]{An observational and theoretical view of the 
radial distribution of HI gas in galaxies}

\author[Jing ~ Wang et al.]{Jing Wang$^{1}$\thanks{Email: wangj@mpa-garching.mpg.de}, Jian Fu$^{2}$, Michael Aumer$^{1,3}$, Guinevere Kauffmann$^{1}$, Gyula I. G. J\'ozsa$^{4,5}$, \\
\newauthor  Paolo Serra$^{6}$, Mei-ling Huang$^{1}$, Jarle Brinchmann$^{7}$, Thijs van der Hulst$^{8}$, Frank Bigiel$^{9}$\\
 $^1$Max--Planck--Institut f\"ur Astrophysik,
        Karl--Schwarzschild--Str. 1, D-85741 Garching, Germany        \\
 $^2$Key Laboratory for Research in Galaxies and Cosmology, Shanghai Astronomical Observatory, CAS, 80 Nandan Rd, Shanghai 200030, China\\
 $^3$Excellence Cluster Universe, Boltzmannstr. 2, D-85748 Garching, Germany\\      
 $^4$Netherlands Institute for Radio Astronomy (ASTRON), Postbus 2, 7990 AA Dwingeloo, The Netherlands \\
 $^5$Argelander-Institut f\"ur Astronomie, Auf dem H.gel 71, D-53121 Bonn, Germany\\
 $^6$Australia Telescope National Facility, CSIRO Astronomy and Space Science, PO box 76, Epping, NSW 1710, Australia\\
 $^7$Leiden Observatory, Leiden University, PO Box 9513, 2300 RA Leiden, The Netherlands \\
 $^8$University of Groningen,  Kapteyn Astronomical Institute, Landleven 12,  9747 AD, Groningen, The Netherlands\\
 $^9$Institut f\"ur theoretische Astrophysik, Zentrum f\''ur Astronomie der Universit\''at Heidelberg, Albert-Ueberle Str. 2, D-69120 Heidelberg\\
 }
\date{Accepted 2014 ???? ??
      Received 2014 ???? ??;
      in original form 2014 January}

\pubyear{2014}
\maketitle

\begin{abstract}

We analyze the radial distribution of HI gas for 23 disk  galaxies
with unusually high HI content from the Bluedisk sample, along with a similar-sized 
sample of "normal" galaxies.
We propose an empirical model to fit the radial profile of the HI surface density,  an 
exponential function with a depression near the center. The radial 
 HI surface density profiles are very homogeneous
in the outer regions of the galaxy; the exponentially declining part of 
the profile has a scale-length of $\sim 0.18$ R1, where R1
is the radius where the column density of the HI is 1 $\ms$ pc$^{-2}$. 
This holds for all galaxies, independent of their stellar or HI mass.
The homogenous outer profiles, combined with the limited range in HI surface density in 
the non-exponential inner disk, results in the well-known tight
relation between HI size and HI mass.
By comparing the radial profiles of the HI-rich
galaxies with those of the control systems, we deduce that 
in about half the galaxies, most of the excess gas lies outside the stellar disk, in the exponentially
declining outer regions of the HI disk. In the other half, the excess
is more centrally peaked.  
We compare our results with existing 
smoothed-particle hydrodynamical simulations and semi-analytic models of
disk galaxy formation in a $\Lambda$ Cold Dark Matter universe.
Both the hydro simulations and the semi-analytic models reproduce the 
HI surface density profiles and the HI size-mass relation 
without further tuning of the simulation and model inputs. 
In the semi-analytic models, the universal shape of the outer HI radial profiles is a
consequence of the {\em assumption} that infalling gas is always distributed
exponentially. The conversion of atomic gas to molecular form explains
the limited range of HI surface densities in the inner disk. These two factors 
produce the tight HI mass-size relation.

\end{abstract}


\section{introduction}
\label{sec:intro}

Over the last three decades, there have been many efforts to map the distribution of HI in galaxies
using radio synthesis telescopes. The first analyses focused on massive and HI-luminous galaxies (e.g. Bosma 1978, Wevers et al. 1984). 
There were also surveys specifically focusing on galaxies in 
clusters (Warmels 1988, Cayatte et al. 1990, Verheijen \& Sancisi 2001, Chung et al. 2009), which demonstrated that  
the HI is depleted as a result of tidal stripping by neighboring galaxies or ram-pressure stripping by the surrounding intracluster medium.                     
WHISP (Westerbork observations of neutral Hydrogen
in Irregular and SPiral galaxies, van der Hulst et al. 2001, Swaters et al. 2002) was one of the largest efforts to map  
the distribution of HI in different types of galaxies. The survey took nearly 10 years, 
observing around 500 galaxies with Hubble types from S0 to Im.   
From these studies, we have learned about the spatial
and kinematic properties of HI in disk galaxies.  The radial HI surface density
profiles exhibit a wide variety of features attributable to irregularities
in the galaxy such as spiral arms, rings, bars, warps etc. Studies of larger samples reveal
basic scaling relations that provide hints of the  mechanisms 
regulating the evolution of galaxies. In contrast to the stellar surface density,
which is peaked in the center of the galaxy and  drops steeply with radius, 
the radial distribution of HI often flattens or even declines  near the center
of the galaxy. In the outer regions, 
the HI disks usually extend to larger radius than the stellar disks and the profiles
are well-fit by exponential functions. 
Studies of different types of galaxies reveal that they all obey the 
same tight HI size-mass relation \citep{Broeils97,Verheijen01,Swaters02,Noordermeer05}. 
Recently, Bigiel \&Blitz (2012) (hereafter B12) found for a sample of 32 nearby spiral galaxies, that the combined HI and $\h2$ gas profiles exhibit 
a universal exponentially declining radial distribution,
if the radius is scaled to R25, the optical radius of the galaxies.

Recently, in an attempt to understand gas accretion in galaxies, we used the Westerbork 
Synthesis Radio Telescope (WSRT) to map hydrogen in a sample of 25 very gas-rich galaxies, 
along with a similar-sized sample of ``control" galaxies with similar masses, 
sizes and redshifts (Wang et al. 2013, Paper I). Thanks to improvements
in the WSRT instrumentation and data analysis, we were able to reach
significantly lower HI column densities  than previous surveys, which
have targeted (late-type) disk galaxies in the field. The sample is selected using
physical properties such as  stellar mass and HI mass fraction,
and is thus well-suited for direct comparisons with theoretical models.
In Paper I, we analyzed the sizes and morphologies of the HI disks. 
The HI-rich galaxies lie on the extension of the well-known HI mass-size relation 
\citep{Broeils97} for normal spiral galaxies, and do not exhibit more
irregular HI morphologies compared to the control galaxies. 
The conclusion is that major mergers are not likely to be the main source of the extra cold gas. 

In this paper, we search for further clues about gas accretion by studying  the radial 
HI surface density profiles of the galaxies in the sample. Again, we compare the    
HI-rich galaxies with the control sample. Our paper is organized as follows: 
section~\ref{sec:data} describes the data used in this paper. In section~\ref{sec:HIprof}, 
we present the radial profiles of HI and propose a model to describe 
the shape of the HI profiles. We present our major observational result that the galaxies exhibit a universal HI radial profile in the outer regions.
In section~\ref{sec:H2prof},  we estimate $\h2$ profiles from the SFR profiles 
for comparison with the results of B12.
We compare our empirical results with 
smoothed-particle hydrodynamics (SPH) simulations and semi-analytic 
models  in section~\ref{sec:simu}. We discuss the possible physical origin
of our main observational results  in
section~\ref{sec:explain}.

\section{Data and previous work}
\label{sec:data}
The Bluedisk project was designed to map the HI in a set of  extremely HI-rich galaxies and
a similar-sized set of control galaxies, which are closely matched in stellar mass,
stellar mass surface density,  inclination and redshift. The aim is to search for clues 
about how galaxy disks grow as a result of gas accretion in the local universe.  
We refer the readers to Paper I for details on the sample selection, 
observations and initial data analysis.   Here, we review the most relevant information. 
All  galaxies that were observed have stellar massed above 10$^{10}$ $\ms$, and redshifts
between 0.01 and 0.03. All galaxies have optical photometry and
spectroscopy from the Sloan Digital Sky Survey \citep[SDSS,][]{Abazajian09} Data Release 7 
spectroscopic MPA/JHU catalog and UV  photometry from the DATA Release 5 of the 
Galaxy Evolution Explorer (GALEX) imaging survey \citep{Martin10}.  We have gathered 
or measured  a variety of parameters \footnote{Please visit http://www.mpa-garching.mpg.de/GASS/Bluedisk/data.php for catalogs of optical and HI properties.} describing the stellar component of the galaxies, 
including sizes (R50 the half-light radius, R90 the radius enclosing 
90\% of the light, and D25 the major axis of the ellipse where the 
$g$-band surface density reaches 25 mag arcsec$^{-2}$ ), stellar mass (M$_*$),
stellar  mass surface density ($\mu_* =0.5M_*/\pi R50^2$), 
concentration index (R90$/$R50), colours and star formation rates (SFR) 
derived from SED (spectral energy distribution) fitting.

Following Paper I, the  sample that will be used in this work includes 42 galaxies. 
5 interacting (multi-source) systems, one strongly stripped system and one non-detection have been
excluded.  The analysis sample is  divided into 23 HI-rich galaxies and 19 
control galaxies, depending on whether the galaxy has more HI mass 
than deduced from its NUV-r color and stellar mass  surface density
using the relation given in Catinella et al. (2010, hereafter C10). 
As in Paper I, our results are derived from the HI total intensity images, 
for which we have produced detection masks and error images. We have measured the 
size R1 (the semi-major axis of the ellipse where the HI surface density 
reaches 1 $\mspc$) for each galaxy. We have also parameterized the 
morphology of the HI disks using  $A$, M20 and Gini from the 
concentration-asymmetry-smoothness (CAS) system \citep{Lotz04,Conselice03},
and 3 newly defined parameters that are more sensitive to irregularities in the outer 
parts of the HI disks, $\Delta$Center, $\Delta$PA and $\Delta$Area. 
The reader is referred to Paper I for more details.

\section{A universal HI profile}
\label{sec:HIprof}
\subsection{Derivation of the angular averaged radial surface density profiles}
\label{sec:deriveprof}
\begin{figure*}
\includegraphics[width=18cm]{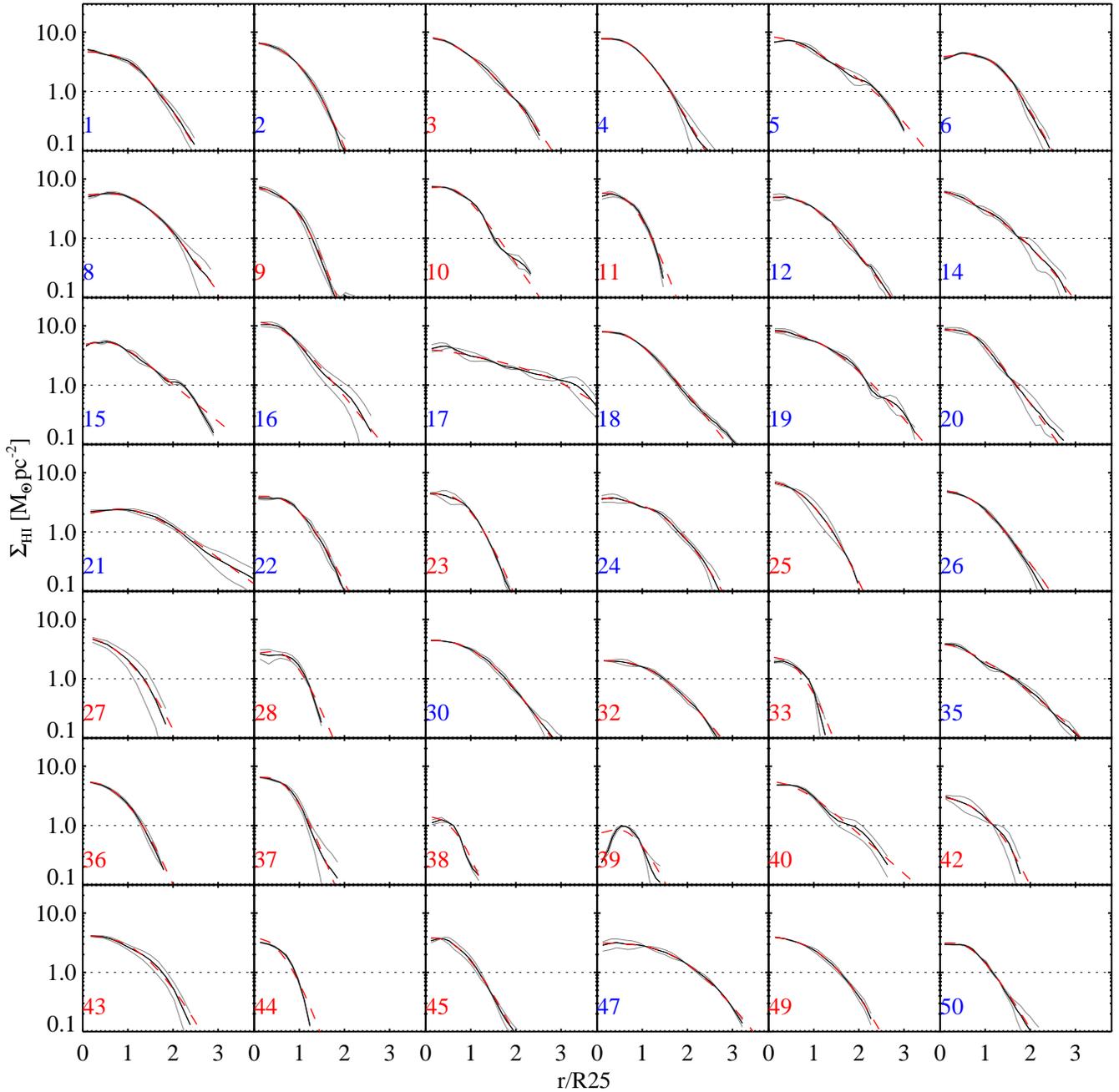}
\caption{Radial profiles of the HI surface density for the 42 galaxies analyzed in this paper. 
The azimuthally averaged profiles are plotted in black, the profiles for
each half of the galaxy divided along the major axis are plotted in grey, 
and the best-fit model fits (see section~\ref{sec:modelHI}) to the profiles
shown in black are plotted as red dashed lines. 
Intersection of the profiles with the dotted lines indicate
where the HI surface density reaches 1 $\mspc$. 
The ID of each galaxy is noted in the corner, coloured  
in blue for HI-rich galaxies and in red for control galaxies. }
 \label{fig:HIprofiles}
\end{figure*}

We measure the angular averaged radial HI surface density profile  along ellipses 
with position angle and axis ratio (a$/$b) determined from the SDSS r-band images. 
The sampling radius (semi-major axis of ellipses) increases linearly with a step 
of 1 pixel (4 arcsec, $\sim$ 2.8 kpc), which is smaller than the PSF, which
is determined by the size of the synthesised beam in our data ($\sim$20 arcsec). 
In Paper I,  we showed that the majority of the galaxies in our sample have
surface density detection thresholds of 0.7$\times 10^{20}$ atoms cm$^{-2}$ at 
a S$/$N (signal-to-noise ratio) of 2. We can go deeper when averaging over a larger area, 
so we can typically measure the radial profiles out to an 
average surface density $\Sigma_{HI}$ of 0.2$\times 10^{20}$ atoms cm$^{-2}$. We do not include the helium gas in the HI-only analysis in this paper. 
The surface densities are corrected to be face-on by multiplying by 
cos$\theta \sim$ b$/$a, where $\theta$ is the inclination angle of the disk. 
We also divide the galaxy into two halves along the major axis and 
measure the radial HI surface density profile  for each half. 
These profiles are displayed as one black and two grey curves 
in Fig.~\ref{fig:HIprofiles}. We can see that the radial profiles for
the two halves of the galaxy  do not usually deviate much 
from the angular-averaged radial profile, for both HI-rich and
control galaxies. In the following sections, we only focus on the angular-averaged HI radial
surface density  profiles (HI radial profiles for short hereafter).   
All the HI radial profiles decline exponentially in the outer regions and flatten or
decline towards the central region, consistent with previous observations of  nearby 
spiral galaxies (e.g. Wevers 1984, Swaters et al. 2002). 

\subsection{Model-fitting}
\label{sec:modelHI}
It has often been remarked that the outer HI surface density profile follows an exponential form beyond the bright, star-forming stellar disk, while in the reach of the disk the surface density is more constant with a possible depression at the center (e.g. Swaters et al. 2002, Bigiel et al. 2012). To describe the shape of a two-component HI radial profile and to obtain the de-convolved shape of it, we choose a simple analytic expression  of the form 
\begin{equation}
\Sigma_{\rm HI,model}(r)=\frac{I_1 \exp(-r/\rs)}{1+I_2*\exp(-r/\rc)},
\label{equ:HIprof}
\end{equation}
and fit our data to it, where $I_1$, $I_2$, $\rs$ and $\rc$ are free parameters. When the radius is large,
the denominator is equal to one and the function reduces to an exponential with scale
radius $\rs$. $\rc$ is the characteristic radius where the profile transitions to the
inner flat region. In this new analytic model, $I_1$ is likely to be connected with the HI$+\h2$ total gas and $I_2$ is likely to be connected with the $I_1/I_2$ atomic-to-molecular gas conversion factor (Leroy et al. 2008).
 As we will show below, this model is very suitable to describe the radial distribution of HI surface density under the resolution of Bluedisk type.

We use the IDL code MPFIT \citep{Markwardt09}, which performs least-squares curve and surface fitting. 
To achieve a fast and stable fit, the whole fitting procedure includes 3 steps.
\begin{enumerate}
\item The first step is to guess the initial value of the model parameters. 
We ignore the angular distribution of HI surface density in the HI image and 
the convolution effect of the beam, and simply perform a linear fit to the 
outer half of the HI radial profile. The slope and intercept distance of 
the linear fit are converted into initial guesses for $r_s$ and $I_1$. 
The initial $\rc$ is set equal to 0.9 $\rs$ and the initial $I_2$ is set equal to  9; 
these are the mean values of these parameters found by Leroy et al. (2008).

\item In the second step, we assume both the HI surface density distribution 
and the beam are azimuthally symmetric. 
With the initial parameters obtained in the last step,  we build a model profile  
and convolve it with one-dimensional Gaussian function with a width of 
FWHM$\sim \sqrt{b_{\rm{min}}\times b_{\rm{maj}}}$, where b$_{\rm{min}}$ and 
b$_{\rm{maj}}$ are the minor and major axis (in FWHM) of the beam. 
The convolved profile ($\Sigma_{\rm{HI,model}}$) is compared with the 
observed profile. The MPFIT code then tunes the model parameters to find the best fit.

\item In the final step, we take into account the axis ratio and position 
angle of both the galaxy and the beam in the HI image. 
The best-fit model parameters from the second step are used as initial input. 
We build a model image with the observed axis ratio and position angle of the galaxy. 
The model image is then convolved with the beam. We measure the radial profiles 
($\Sigma_{\rm{HI,model}}$) from the convolved model image in the
same way as they are measured from the observed HI image (Sect. 3.1).
The simulated profiles are compared to the observed HI radial profile. 
The MPFIT code is then used to tune the model parameters to  find the best-fit.
\end{enumerate}
The best-fit model profiles are displayed as red dashed lines in Fig.~\ref{fig:HIprofiles}. 
We can see that in general the model describes the observed profiles very well.

\begin{figure*}
\includegraphics[width=16.cm]{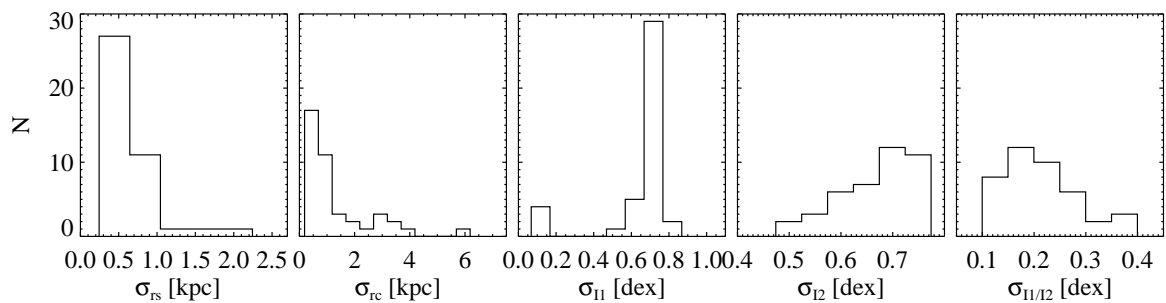}
\caption{Distribution of errors on the parameters,$\rs$, $\rc$, $I_1$,  $I_2$ and $I_1/I_2$  in the best-fit models for the HI radial profiles.}
 \label{fig:fiterror}
\end{figure*}

We now estimate errors for our derived parameters, as well degeneracies in the fits. 
We construct a grid around the best-fit parameters: $\rs$ and $\rc$ vary 
with a step size of 0.02 R25  in a range $\pm$R25 about the best-fit
values, while and $I_1$ and $I_2$ vary 
with step size of 0.01 dex in a range $\pm$1 dex about the best values. 
We calculate the sum of square residuals between the convolved model and the data at each grid point
\begin{equation}
\kappa^2=sum(\frac{(\Sigma_{\rm{HI,model}}-\SHI)^2}{\sigma(\SHI)^2}), 
\end{equation}
 and take $p_i=$exp(-$\kappa^2/2$) as the goodness-of-fit of the 
parameters for modelling the data at the grid point $i$. 
We thus build a probability distribution function (PDF) within the parameter grids 
by assigning  $p_i$ as the relative probability of the data to be modelled by the 
parameters at grid $i$. We project the PDF in the direction of each parameter, 
and find the contours that contain  99.5 percent of the 
cumulative probability distribution.  The half-width of these contours correspond to  
 3 times $\sigma$ (error) of the parameter. 
The derived errors for  the whole sample are listed in Table~1 
(and displayed as a histogram in Fig.~\ref{fig:fiterror}). We summarize our
results as follows: 
\begin{enumerate}
\item $\rs$ is well constrained ($\sigma<$ 1 kpc) for most of the galaxies in the Bluedisk sample.
\item $\rc$ often has large error bars ($\sigma>$2 kpc),  due to the resolution limit of the HI images.
\item There is strong degeneracy between $I_1$ and $I_2$ for all the galaxies, 
the 1 $\sigma$ errors for both $I_1$ and $I_2$ are larger than 0.3 dex.
\item $I_1/I_2$ can be well-constrained with $\sigma\sim$0.2 dex for nearly half of the galaxies, but when the HI disks are relatively small (R$_1<$30 arcsec), 
$I_1/I_2$ cannot be robustly determined and has large errors. 
\end{enumerate}

\begin{figure}
\includegraphics[width=7cm]{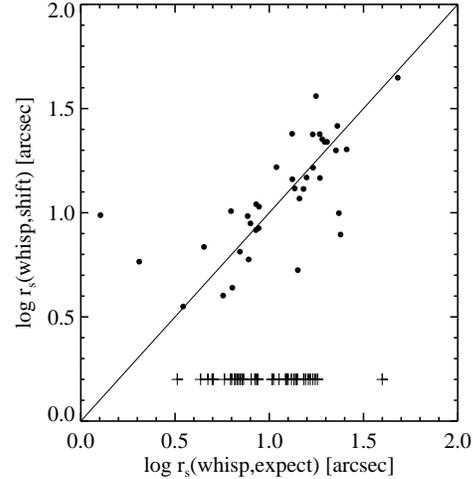}
\caption{WHISP galaxies are shifted and convolved with the WSRT beam to have similar 
appearance to the Bluedisk galaxies. $\rs$(shift) are the scale-length measured from 
the shifted and convolved images, and $\rs$(expected) are the true scale-length expected 
at the redshifts of the Bluedisk galaxies. The crosses show the sizes of the Bluedisk galaxies.} 
 \label{fig:shiftwhisp}
\end{figure}

We test our method with galaxies from the WHISP survey 
\citep{Swaters02}   
to ensure that our estimates of $\rs$
will not be compromised by the  resolution of Bluedisk HI images. We select the galaxies with optical images 
available from the SDSS and with stellar masses greater than 10$^{9.8}$ M$_{\odot}$. 
The publicly available total-intensity maps of the WHISP galaxies (with a median distance of z$\sim$0.008) are rescaled in angular size
so that they have similar dimensions to the Bluedisk galaxies (with median distance of 
z$\sim 0.026$), and convolved with a Gaussian PSF of 22$\times$16 arcsec$^2$.  
We fit the model to the HI profiles from both the original WHISP  images and the rescaled images. 
We compare these two estimates of  $\rs$ in Fig.~\ref{fig:shiftwhisp}. 
The two estimates  correlate with a correlation coefficient of 0.7 and there
is no systematic offset. The scatter around the 1-to-1 line is 3.3 kpc, larger than the error listed in Table 1, because the WHISP data is much shallower than the Bluedisk data.

\begin{table*}
\begin{center}
\renewcommand{\arraystretch}{1.}
\begin{tabular}{ccccccccc}
\hline \hline
 \multicolumn{1}{c}{ID} & \multicolumn{1}{c}{$\rs$} & \multicolumn{1}{c}{$\sigma_{rs}$} & \multicolumn{1}{c}{$\rc$} & \multicolumn{1}{c}{$\sigma_{rc}$} & \multicolumn{1}{c}{$I_1$}   & \multicolumn{1}{c}{$\sigma_{I1}$} & \multicolumn{1}{c}{$I_2$}  & \multicolumn{1}{c}{$\sigma_{I2}$} \\

 \multicolumn{1}{c}{ } &   \multicolumn{1}{c}{kpc}   & \multicolumn{1}{c}{kpc} & \multicolumn{1}{c}{kpc} & \multicolumn{1}{c}{kpc} & \multicolumn{1}{c}{$\mspc$} &  \multicolumn{1}{c}{dex} & \multicolumn{1}{c}{$\mspc$} &  \multicolumn{1}{c}{dex}   \\
\hline
   1&      6.91&      0.64&      6.35&      0.74&    181.15&      0.68&     23.80&      0.61\\
   2&      3.95&      0.40&      3.91&      0.51&    760.45&      0.64&     35.73&      0.57\\
   3&      4.69&      0.41&      7.02&      1.31&    335.88&      0.72&     13.25&      0.68\\
   4&      7.20&      0.61&      3.58&      0.55&    113.87&      0.62&     11.18&      0.52\\
   5&      6.17&      0.98&      8.78&      2.97&   1162.44&      0.69&     79.04&      0.65\\
   6&      7.83&      0.60&      6.32&      0.40&    310.51&      0.80&     45.73&      0.77\\
   8&      7.84&      0.74&      5.79&      0.75&     64.62&      0.67&      9.94&      0.55\\
   9&      3.62&      0.38&      3.28&      0.73&   1060.26&      0.69&     65.03&      0.63\\
  10&      6.53&      0.66&      2.54&      1.14&     26.31&      0.68&      7.33&      0.72\\
  11&      4.23&      0.36&      2.75&      0.48&    461.41&      0.76&     84.26&      0.72\\
  12&      8.50&      0.93&      5.69&      2.21&     72.57&      0.68&      6.10&      0.67\\
  14&      8.27&      1.21&     12.14&      6.09&    245.10&      0.60&     11.96&      0.73\\
  15&     22.32&      1.63&      7.48&      2.82&     14.66&      0.57&      2.34&      0.47\\
  16&      5.86&      0.39&      0.66&      0.19&     31.13&      0.16&     43.32&      0.76\\
  17&      7.71&      2.24&      9.36&      0.25&    476.54&      0.11&     62.71&      0.76\\
  18&      8.52&      0.63&      1.48&      1.10&     68.59&      0.66&     60.54&      0.59\\
  19&      7.47&      1.00&      7.45&      3.60&    124.83&      0.69&      9.22&      0.68\\
  20&      7.15&      0.47&      1.74&      0.69&     48.02&      0.59&     17.51&      0.49\\
  21&      9.10&      0.84&      1.95&      1.08&     20.93&      0.68&     63.86&      0.70\\
  22&      5.45&      0.51&      3.13&      0.52&    305.92&      0.70&     47.68&      0.65\\
  23&      5.12&      0.45&      3.59&      0.87&    365.23&      0.72&     53.47&      0.69\\
  24&      7.46&      0.61&      7.74&      0.81&    230.00&      0.73&     24.98&      0.70\\
  25&      4.11&      0.58&      4.56&      3.79&    224.27&      0.69&     16.37&      0.66\\
  26&      2.99&      0.24&      2.05&      0.41&    540.47&      0.73&     43.68&      0.73\\
  27&      1.85&      0.45&      1.26&      2.15&    385.18&      0.75&     67.09&      0.62\\
  28&      3.32&      0.35&      2.01&      0.40&     52.62&      0.70&     33.62&      0.76\\
  30&      6.51&      0.64&      4.31&      1.03&    440.29&      0.69&     34.81&      0.62\\
  32&      3.12&      0.25&      2.28&      0.43&    272.01&      0.70&     75.07&      0.68\\
  33&      2.95&      0.54&      1.79&      0.54&     81.20&      0.70&     38.89&      0.75\\
  35&      7.92&      0.66&     14.15&      0.23&    579.77&      0.07&     23.17&      0.77\\
  36&      2.45&      0.40&      1.62&      2.00&    334.37&      0.71&     60.83&      0.72\\
  37&      3.53&      0.39&      1.93&      0.52&    147.55&      0.67&     38.81&      0.71\\
  38&      2.69&      0.78&      1.44&      3.00&     38.32&      0.78&     50.66&      0.75\\
  40&      7.56&      0.74&      0.47&      0.21&     13.43&      0.13&     66.24&      0.76\\
  42&      3.53&      0.78&      3.80&      3.33&    216.74&      0.76&     29.05&      0.59\\
  43&      3.18&      0.32&      2.77&      0.74&    332.24&      0.70&     57.64&      0.66\\
  44&      2.24&      0.30&      1.47&      0.35&    268.87&      0.75&     59.20&      0.77\\
  45&      3.40&      0.32&      1.97&      0.82&    334.05&      0.70&     64.08&      0.64\\
  47&      6.50&      0.91&      6.25&      1.38&    318.16&      0.69&     53.31&      0.62\\
  49&      3.80&      0.37&      3.77&      0.66&    688.92&      0.72&     74.17&      0.69\\
  50&      5.18&      0.57&      3.02&      1.32&    464.34&      0.69&     73.66&      0.63\\
\hline \hline
\end{tabular}
\caption{Parameters of the best-fit models for the HI radial profiles.}
\label{tab:fitHI}
\end{center}
\end{table*}

\subsection{The universal HI radial profiles in outer disks}
\label{sec:uniprof}
\begin{figure*}
\includegraphics[width=14cm]{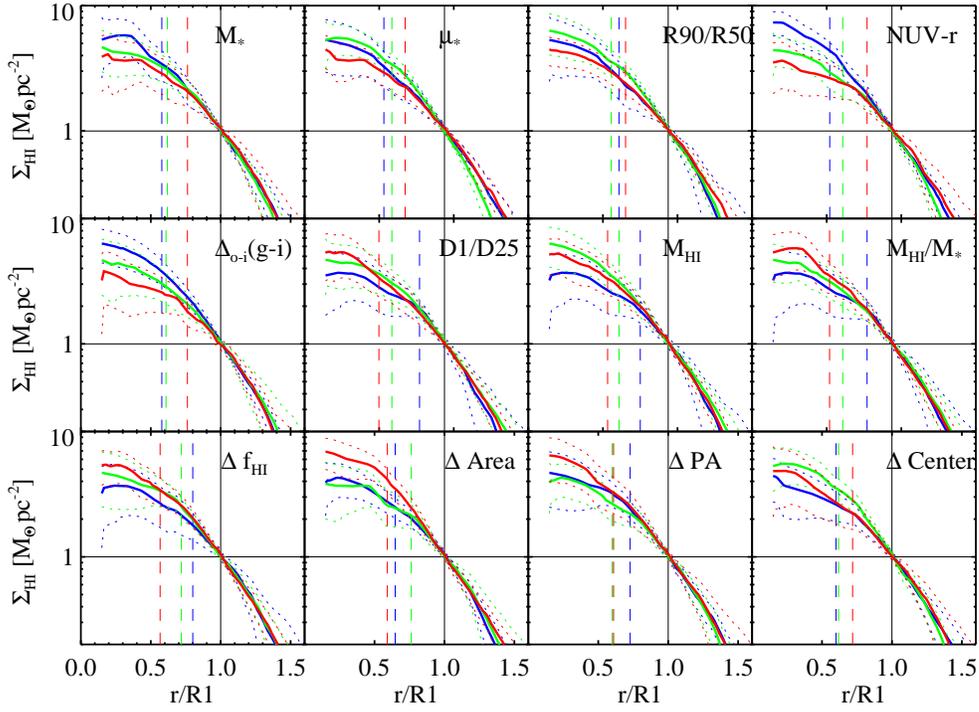}
\caption{The galaxy sample is equally divided into 3 sub-samples according to 
the parameter denoted at the top right corner of each plot. These include stellar mass (M$_*$), 
stellar mass surface density ($\mu_*$),  concentration index (R90$/$R50), NUV--$r$ colour, 
colour gradient ($\Delta_{o-i}(g-i)$), HI-to-optical size ratio (D1$/$D25), 
HI mass ($\mHI$), HI mass fraction ($\HIs$), HI excess ($\dfHI$), and HI 
morphological parameter $\Delta$Area, $\Delta$PA, and $\Delta$Center. 
The median HI profile for each  sub-sample is plotted in blue for
the lowest values of each parameter, in red for the  highest values and
in green for intermediate values. The dotted lines show the scatter arround the median profiles.
 The dashed lines show the median value of R25$/$R1 for
each of the sub-samples . The solid lines indicate R$1=1$ and $\SHI=1 \mspc$.
}
 \label{fig:avgprof_r1}
\end{figure*}

We divide the Bluedisk sample into 3 equal sub-samples by their stellar properties 
(stellar mass, stellar mass surface density, colour, and colour gradient) and HI properties 
(HI:optical size ratio, HI mass fraction, HI mass, deviation from the C10 plane (,see section~\ref{sec:data}), and
the morphological parameters $\Delta$Area and $\Delta$Center). Instead of scaling the profiles
by R25, the characteristic {\em optical} radius of the disk, we
scale by R1, the radius where the column density of the HI is 1$\ms$ pc$^{-2}$.  
The median HI radial profiles of the sub-samples are displayed in Fig.~\ref{fig:avgprof_r1}. 
The outer HI profiles now display ``Universal'' behaviour and exhibit an 
exponentially declining profile  from $\sim$0.75 R1  to 1.3R1. 
Even the HI-rich galaxies (with large $\dfHI$), which were postulated by C10
to have experienced recent gas accretion, the clumpy HI disks with large $\Delta$Area, 
or HI disk that are off-center with respect to the optical disk (with large $\Delta$Center), 
have outer HI profiles that do not deviate significantly. We
note that the  median value of the  
averaged surface between 0.75 R1 and 1.25 R1 is  $\sim1.142\pm0.022\mspc$.

The differences between the inner profiles (at radii smaller than 0.75 R1) are much more prominent.
They are most significant when the sample is split by NUV-r colour.  
In the next sections, we will show that in semi-analytic models that treat the conversion of atomic gas
into molecular gas, differences in the inner HI surface densities arise because the gas reaches
high enough surface densities in the central region of the galaxy to be transformed into
molecular form. Because the level of star formation activity in the galaxy is tied to
its molecular gas content, correlations between star formation rate
and HI surface density will arise naturally.

\begin{figure*}
\includegraphics[width=14cm]{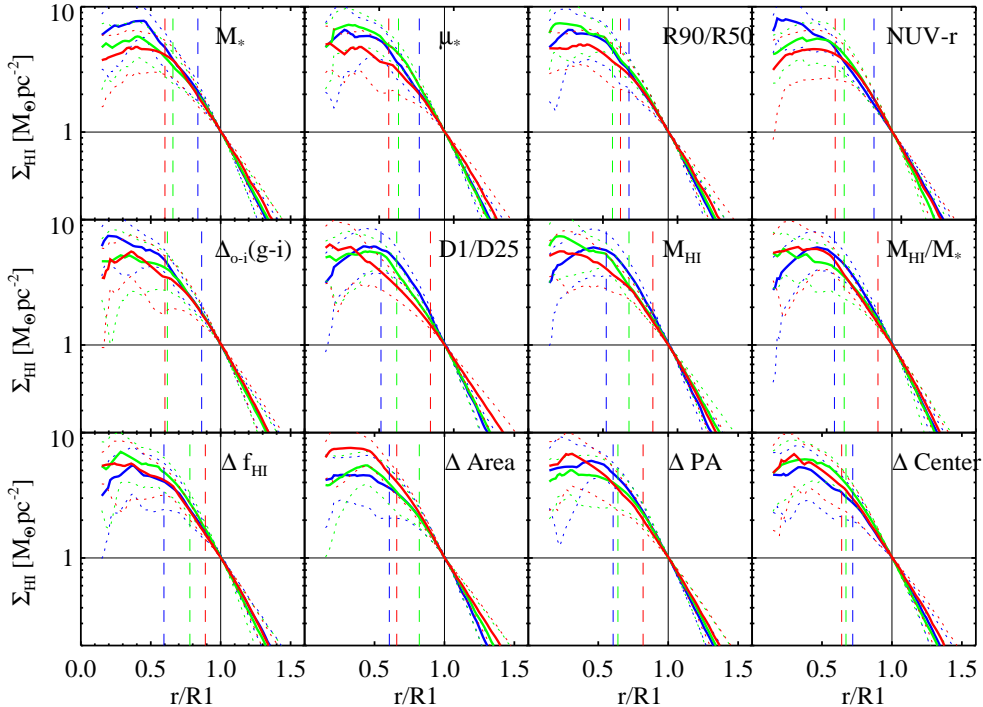}
\caption{Similar as Fig.~\ref{fig:avgprof_r1}, but for the  best-fit model profiles.}
 \label{fig:avgprof_r1a}
\end{figure*}

\begin{figure*}
\includegraphics[width=14cm]{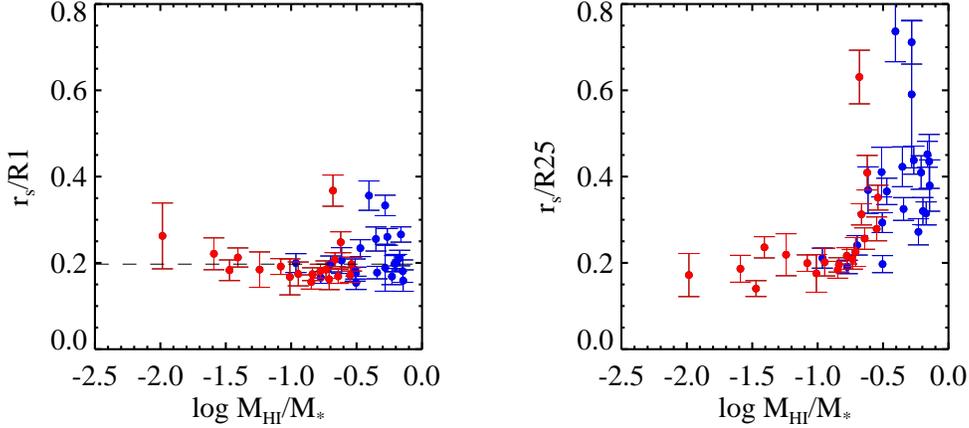}
\caption{$\rs$/R1 (left) and $\rs$/R25 (right) are plotted as a function of HI mass fraction.
The black dashed line shows the median value of $\rs/$R1. HI-rich galaxies are plotted in blue,
while control galaxies are plotted in red. }
 \label{fig:Rs_R1}
\end{figure*}

In Fig.~\ref{fig:avgprof_r1a}, we plot the best-fit model profiles rather than the actual profiles. 
We can see that the results in the outer regions of the disks do not change.  
The median  $\Sigma_{\rm{HI,0.75-1.25R1}}$ from the best-fit models  is  $\sim$1.083$\pm$0.063 $\mspc$.  
We plot the ratios  $\rs$/R1 and $\rs$/R25 as a function of HI mass fraction in   
Fig.~\ref{fig:Rs_R1}.  From the left panel, we can see that 
the dispersion of $\rs$/R1 around the median value of 0.19$\pm$0.006 is small. 
 There are 2 HI-rich galaxies (galaxy 15 and 21) and one control galaxy (galaxy 40) 
which deviate from the mean $\rs/$R1 by more than 0.1. These galaxies were not peculiar in terms of their
HI mass (fractions), morphologies,  global profiles, nor did their tellar properties differ in any clear way from
the other galaxies in the sample.

From the  right panel of Fig.~\ref{fig:Rs_R1},  we can seen that $\rs/$R25
spans a much wider range of values, from 0.15 to 0.8, with most of the values below 0.6. 
The HI-rich galaxies have systematically higher HI-to-optical size ratios and also
span a larger range in values.

\subsubsection{Relation to the HI mass-size relation} \label{sec:HIMD}

It is well known that in disk galaxies, the HI mass  and the HI size R1 are tightly
correlated \citep[][ and Paper I]{Broeils97,Verheijen01,Swaters02,Noordermeer05}. 
\citep{Broeils97} parametrize this relation as
\begin{equation}
\frac{{{M_{{\rm{HI}}}}}}{{{M_ \odot }}} = \left( {12.88 \pm 3.85} \right){\left[ {\frac{{R1}}{{{\rm{pc}}}}} \right]^2}
 \label{equ:eq1}
\end{equation}

\begin{figure*}
\includegraphics[width=14cm]{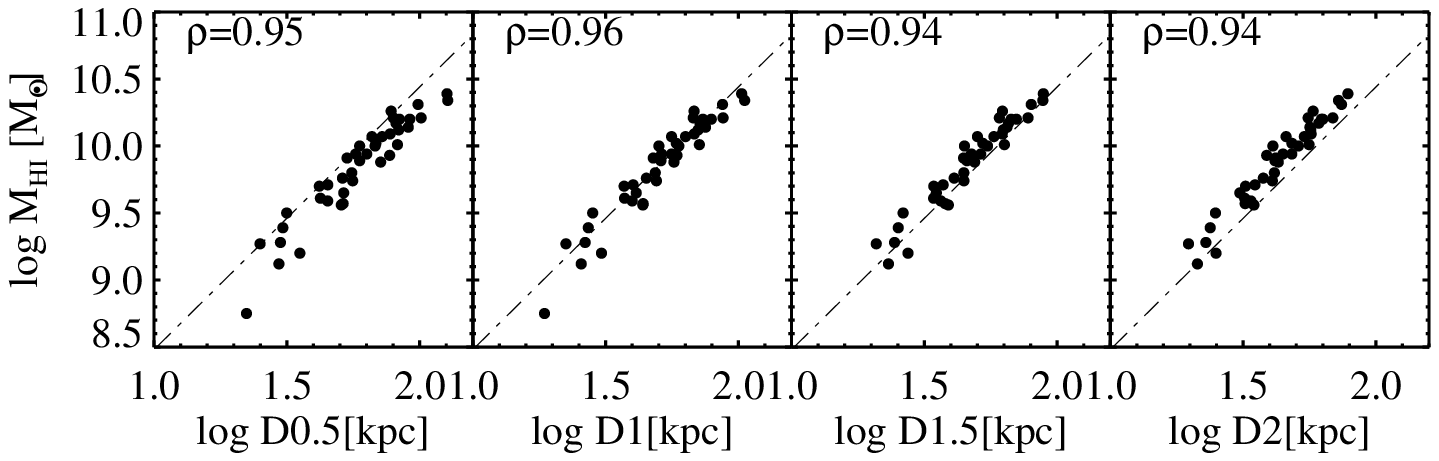}
\caption{ The relation between HI mass and the diameters of HI disks measured at a surface density of 0.5, 1, 1.5 and 2 $\mspc$. The correlation coefficient $\rho$ is denoted in the corner of each figure. The dashed line shows the HI mass-D1 relation from Broeils et al.(1997).}
 \label{fig:D_MHI}
\end{figure*}

Our finding that the HI gas in the outer region of disks has a "universal" exponential surface density 
profile offers a natural explanation for the tight HI mass-size relation.
Our results also suggest that similar tight relations should be found for
different definitions of HI size. This is illustrated in Fig.~\ref{fig:D_MHI}, where we show the 
correlations between $\mHI$ and the radii where the HI surface density 
reaches 0.5, 1,  1.5 and 2 $\mspc$, respectively. As can be seen,  
the correlations are equally strong and tight for all four radii. The scatter in the 
HI mass-size relation only increases significantly at radii coreesponding to surface densities 
greater than a few $\mspc$, where the total cold gas content is expected to
be dominated by molecular rather than atomic gas.

\subsubsection{Where is the ``excess'' HI gas located?}
\label{sec:excessHI}
\begin{figure}
\includegraphics[width=7cm]{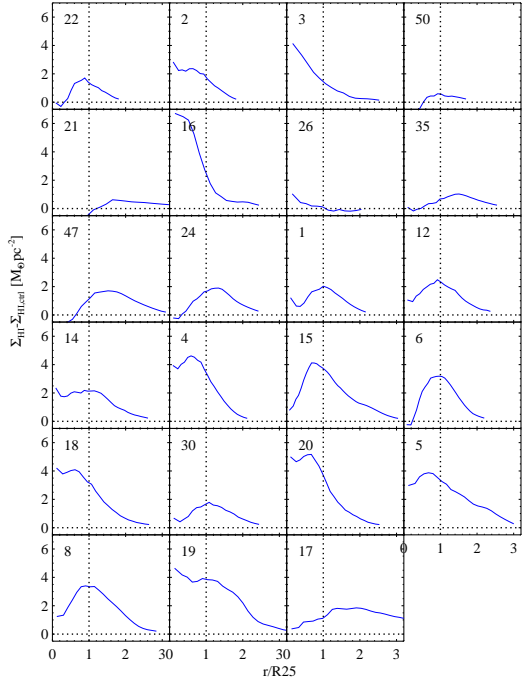}
\caption{The difference between the HI profiles for the HI-rich galaxies 
and the median HI profile of the control galaxies. 
The dashed lines show where the radius equals R25. The IDs of the HI-rich galaxies 
are denoted in the corner of each panel.}
 \label{fig:extraHI}
\end{figure}

In this section, we subtract the median  HI profile of galaxies in the ``control'' sample from
the HI profile of each HI-rich galaxy, in order to study the spatial distribution of the
excess gas. Results are shown in 
Fig.~\ref{fig:extraHI}. Interestingly, there appear to be one  class
of galaxy (e.g. 2,3, 16, 14, 4, 18,20,5, 19)
where the excess rises montonically towards the smallest radii, or peaks well within the optical radius R25.
In a second class of galaxy (e.g. 22, 35, 47, 24,1 12, 15, 6, 30, 8, 17), the excess peaks
very close to R25 and drops off on either side. 
The latter is suggestive of {\em recent accretion of a ring of gas
at the edge of the optical disk.}

\section{The total cold gas profiles}
\label{sec:H2prof}
B12 analyzed average gas radial distributions of 17 star-forming massive disk galaxies. They found that 
if they re-normalized the surface densities by dividing by the transition 
surface density $\Sigma_{\rm{transit}}$ where $\Sh2$ equals $\SHI$,
and dividing the radii by R25, the characteristic {\em optical} radius of the disk,
the resulting gas radial profiles exhibited a ``Universal'' exponential distribution beyond 0.2 R25,
which could be parameterized as:
\begin{equation}
\Sigma_{\rm{gas}}/\Sigma_{\rm{transit}}=2.1 e^{-1.65 \rm{r/R25}}
\label{equ:b12}
\end{equation}

In this section we explain how we construct total cold gas radial profiles for our Bluedisk galaxies,
for comparison with the B12 results. We do not have information 
about the molecular gas in  all the Bluedisk galaxies, and the spatial resolution of the 
Bluedisk HI data (FWHM of PSF $\sim$ 10.6 kpc) is lower than the 
B12 data (FWHM of PSF $\sim$ 1 kpc).  

First, we use the star formation rate radial profiles to provide an estimate of 
the $\h2$ radial profiles for our galaxies. We use the GALEX FUV images and the 
WISE 22 $\mu$m images to derive the SFR profiles ($\Sigma_{\rm{SFR}}$). 
The FUV images were converted to the resolution of the 22 $\mu$m images. 
Following Wright et al. (2010) and Jarrett et al. (2011),  we
apply a colour correction to the  22 $\mu$m luminosity 
of extremely red sources. We  calculate the IR-contribution to the
star formation rate (SFR$_{IR}$) of our sample by adopting the relation 
between SFR and WISE 22 $\mu$m luminosity in Jarrett et al. (2013). We adopt
the formula from Salim et al. (2007) to derive the UV contribution to
the star formation rate,  SFR$_{FUV}$. The final SFR is the sum of 
SFR$_{IR}$ and SFR$_{FUV}$.
$\Sigma_{\rm{SFR}}$ is then converted into an estimated $\h2$ surface density 
using the formula in Bigiel et al. (2008).
 The total gas radial profile is calculated as
\begin{equation}
\Sigma_{\rm gas}=1.36 (\Sigma'_{\h2}+\Sigma_{\rm HI}),
\end{equation}
where the factor of 1.36 corrects for the contribution from helium. 

\begin{figure}
\includegraphics[width=9cm]{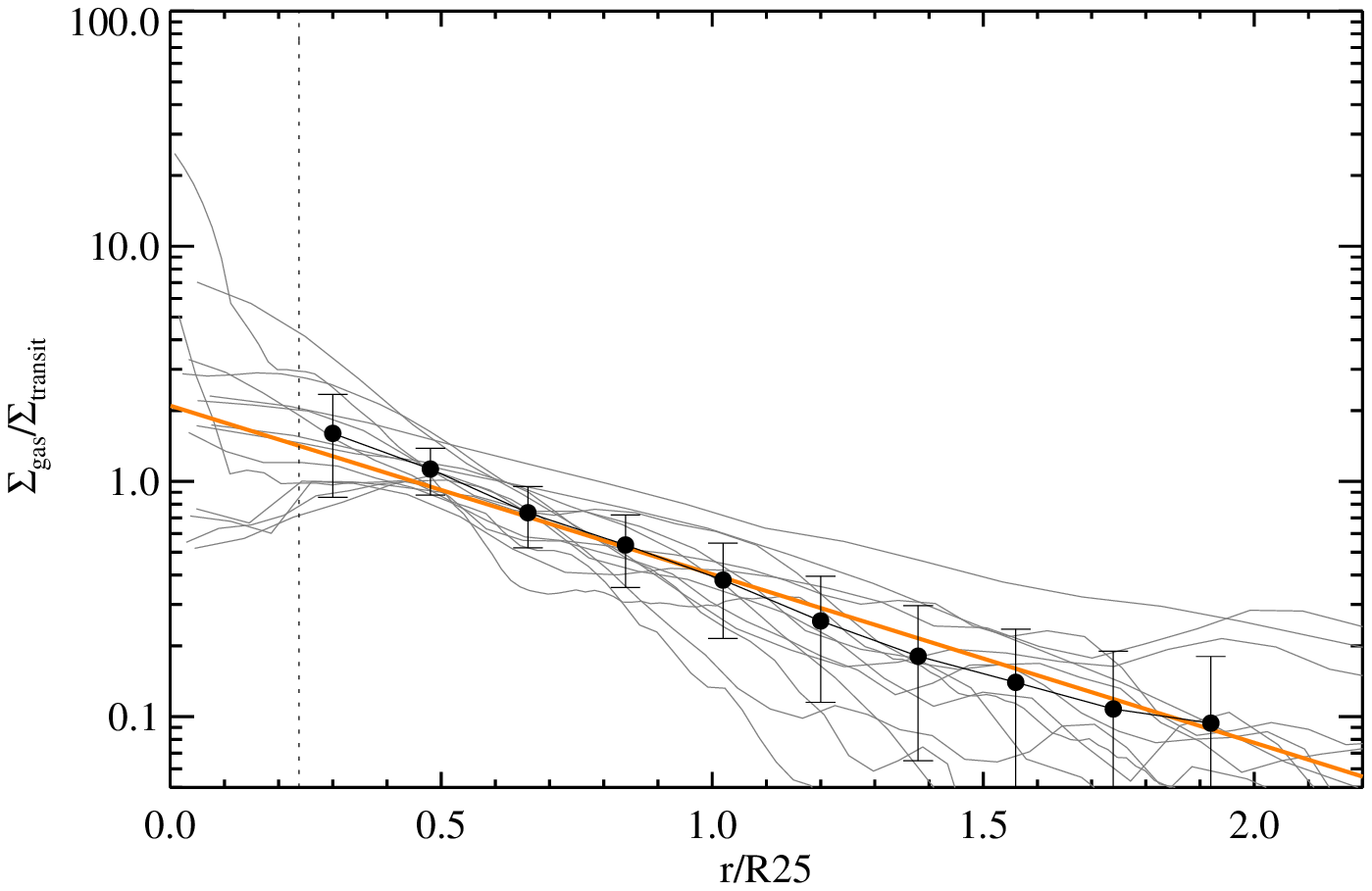}
\includegraphics[width=9cm]{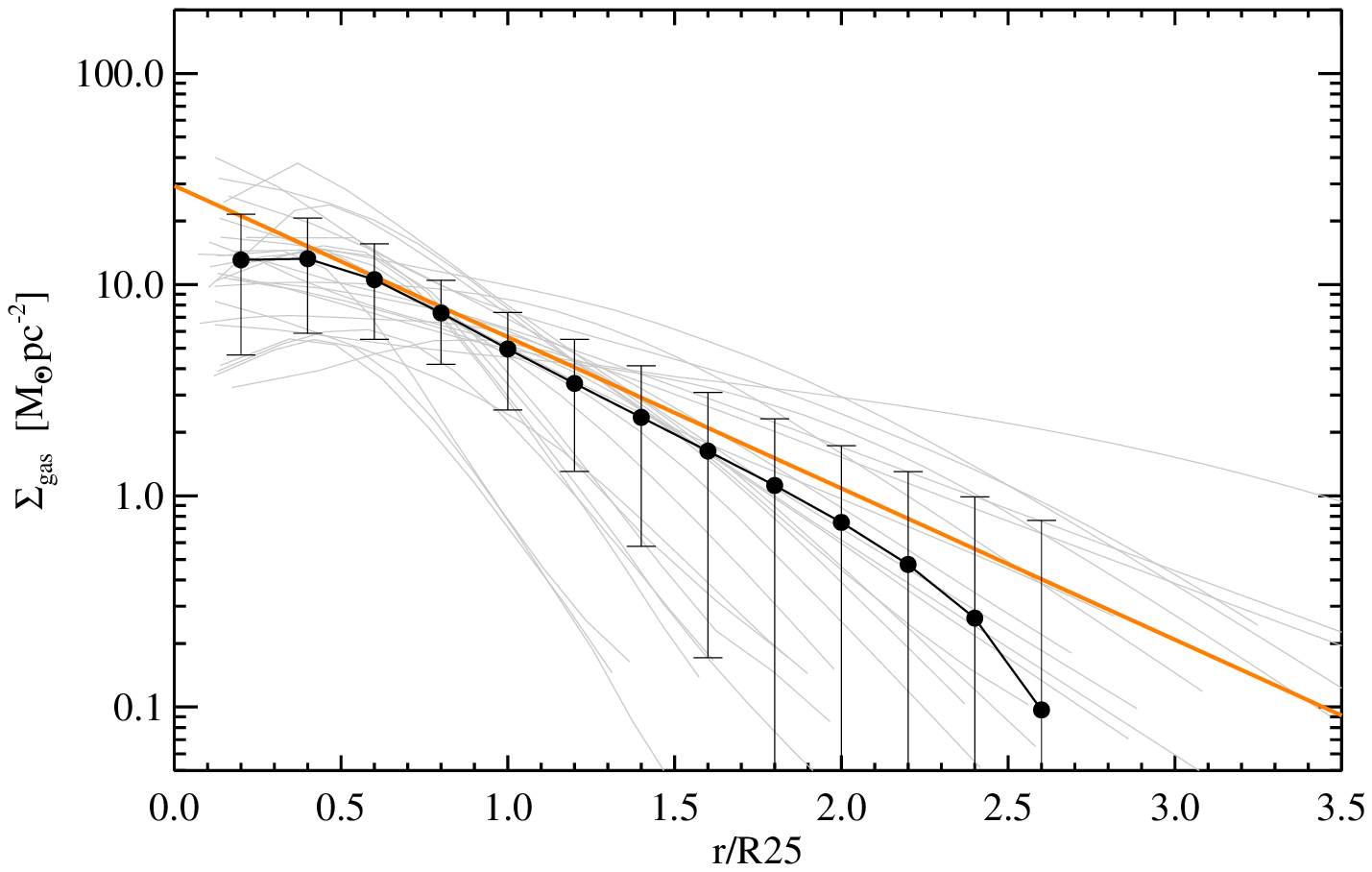}
\includegraphics[width=9cm]{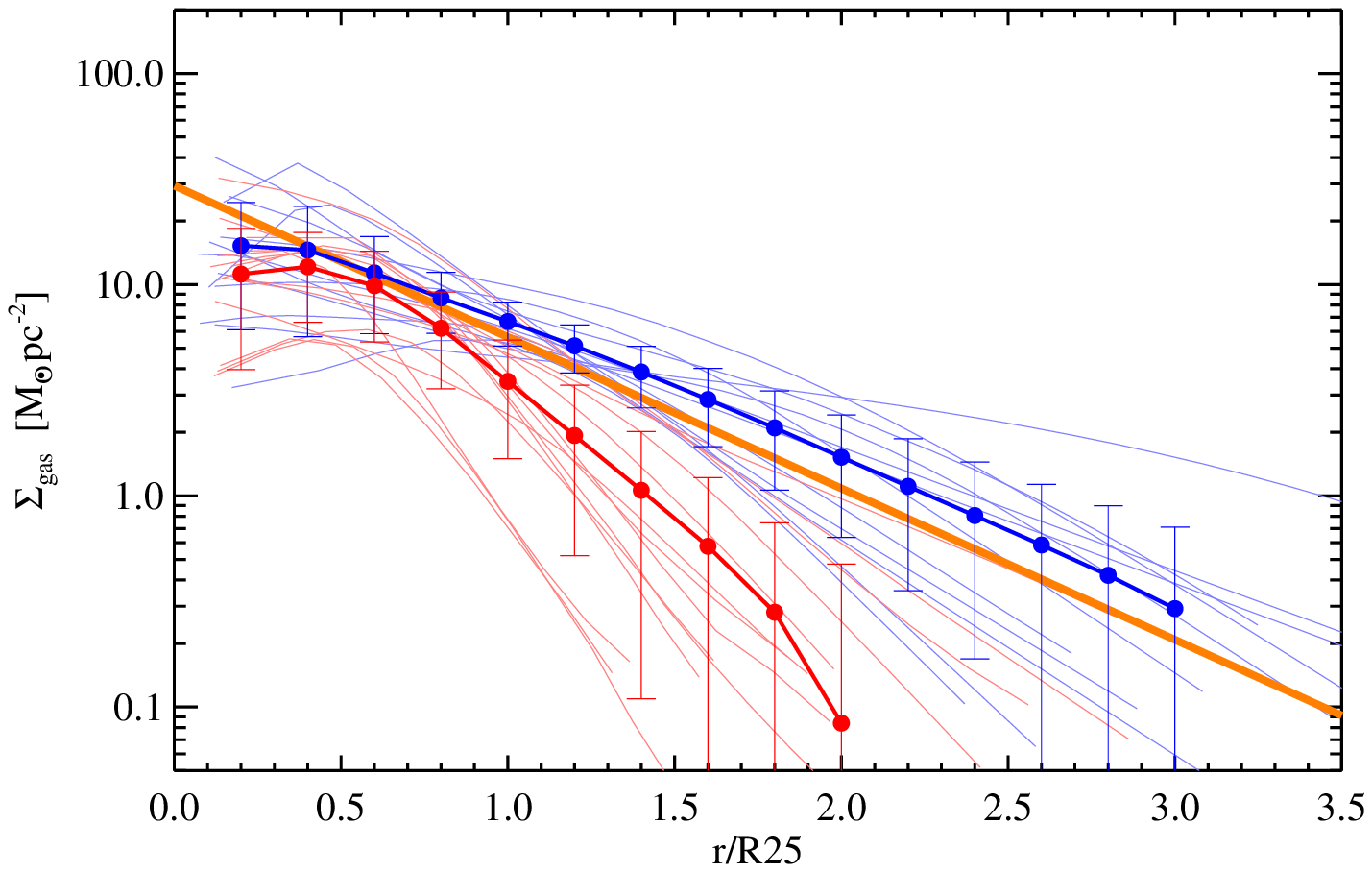}
\caption{
Top panel: estimated total gas profiles for the WHISP galaxies (grey). The black points show
the average profile and the  error bars indicate the 1$\sigma$ scatter derived from boot-strapping. 
The orange line shows the universal relation obtained by B12 when interacting systems are excluded. 
Middle panel:  similar to the top panel, but for the Bluedisk galaxies. A fixed transition 
surface density $\Sigma_{\rm{transit}}\sim$ 14 $\mspc$ is assumed (Leroy et al. 2008). 
Bottom panel: The same as the middle panel, except that the red lines are for the 
control galaxies and the blue lines are for the HI-rich galaxies.} 
 \label{fig:tgasprof}
\end{figure}

To confirm that this technique is robust, we apply it to the sub-sample of  galaxies 
from the WHISP survey, which have GALEX NUV and  WISE 22 $\mu$m images. 
Following the sample definition in B12, we select
massive (M$_*/\ms>10^{9.8}$), relatively face-on (size ratio b$/$a$>$0.5, or inclination angle smaller than 60 deg) star forming (NUV--$r<$3.5) disk (concentration $R90/R50<2.8$) galaxies.
The WISE 22 $\mu$m images have similar resolution to the B12 images. 
The HI images of the WHISP galaxies  have a FWHM PSF of $\sim$2 kpc, again
very similar to that of the B12 images. 
The black points in the top panel of Fig.~\ref{fig:tgasprof} show the 
average, scaled cold gas radial surface density profile. The relation derived by B12 is shown in orange, and
the agreement is very good. 

Following B12, to minimize resolution effects, the analysis is performed on the relatively face-on 
(b$/$a$>$0.5) Bluedisk galaxies, which include 12 HI-rich galaxies and 14 control galaxies.
We note that the HI maps of the Bluedisk galaxies have considerably worse resolution 
(FWHM of PSF $\sim$ 10 kpc) than the B12 gas maps.  We use the  best-fit model HI 
profiles derived in the last section, instead of the directly-measured HI radial profiles. 
Because $\Sigma_{\rm{transit}}$ varies little from galaxy to galaxy, we assume a fixed 
$\Sigma_{\rm{transit}}\sim$ 14 $\mspc$ from the B12 relation (Leroy et al. 2008) 
rather than calculating $\Sigma_{\rm{transit}}$ from the best-fit model HI profile.
This is because the fit in the inner region has large uncertainties (section \ref{sec:modelHI}). 
Our results are shown in the middle and bottom panels of Fig.~\ref{fig:tgasprof}. 
From the middle panel of Fig.~\ref{fig:tgasprof}, we see that within 2R25 the mean gas profile 
of the Bluedisk sample is also consistent with the B12 relation. 
The individual profiles exhibit a large scatter around the B12 relation in the outer regions (beyond R25). 
In the bottom panel of Fig.~\ref{fig:tgasprof}, we plot the mean gas profile of the control
galaxies in red and the mean profile of the HI-rich galaxies in blue.  
The mean profile of the HI-rich galaxies lies slightly above the B12 relation, 
while that of the control galaxies lies significantly below the B12 relation. 
This is consistent with our previous findings (Paper I) that HI-rich galaxies have larger 
R1$/$R25 than the control galaxies.

To summarize , the radial gas profile of Bluedisk galaxies exhibit large scatter around 
the B12 relation in their outer regions. The control galaxies deviate significantly from the B12 relation, which may be connected with the low accretion rate of cold gas or more evolved HI disks.

\section{Comparison with SPH simulations and Semi-Analytic Models} \label{sec:simu}
In this section, we will compare the results presented in the previous
sections with the SPH simulations of Aumer et al. (2013) and the semi-analytic
models of  Fu et al. (2013).

\subsection{Comparison with SPH simulations} \label{sec:SPH}

We compare our observed HI profiles to those found in 19 $z=0$ galaxies from cosmological zoom-in SPH simulations.
The simulations were described in detail in \citet{a13} and follow the formation of galaxies in haloes with
masses $1\times 10^{11}<M_{200}/M_{\odot}<3\times 10^{12}$ in a $\Lambda$ Cold Dark Matter universe.
The simulations include models for multiphase gas treatment, star formation, metal enrichment, metal-line cooling,
turbulent metal diffusion, thermal and kinetic supernova feedback and radiation pressure from young stars.
They lack, however, a model for the partition of atomic and molecular gas in interstellar medium.

As was shown in \citet{a13}, the $z=0$ model galaxies all contain gas disks, 
with gas fractions that are  higher
than the observed average at the same stellar mass. The model disk galaxies
are thus quite comparable in HI content to the  Bluedisk galaxies.
To obtain HI profiles for the simulated disks, we apply a simple model based on the observationally motivated
model by \citet{blitz}. The same model was used in the semi-analytic models of \citet{fu}.

We first select all gas with $T<15000\;\rm{K}$ and determine its angular momentum vector to orient
the galaxy in a face-on orientation. We then divide the galaxies into quadratic 
pixels with $500 \;\rm{pc}$ side-length.
According to \citet{blitz} the ratio of molecular and atomic hydrogen is
\begin{equation}
f_{\rm mol} (x,y)= \Sigma_{\rm H_2}(x,y)/\Sigma_{\rm HI}(x,y) = \left(P(x,y)/P_0\right)^{\alpha},
\end{equation}
where $\alpha=0.92$ and $P_0=5.93\times10^{-13}\;{\rm Pa}$ are constants fit to observations.
For the mid-plane pressure $P(x,y)$ we use the relation given by \citet{elmegreen},
\begin{equation}
P(x,y)={{\pi}\over{2}}G\left[\Sigma_{\rm gas}(x,y)^2+f_{\sigma}(x,y)\Sigma_{\rm gas}(x,y)\Sigma_{\rm stars}(x,y)\right]
\end{equation}
, with the ratio of gas-to-stellar velocity dispersions
\begin{equation}
f_{\sigma}(x,y)={{\sigma_{\rm gas}(x,y)}\over{\sigma_{\rm stars}(x,y)}},
\end{equation}
which, as the surface densities, we take directly from the simulations.
To account for ionized hydrogen among the $T<15000\;\rm{K}$ gas, we assume a 25\% ionized fraction in stellar
dominated regions and correct for the absence of stars as a source of ionization by crudely invoking a factor
$\Sigma_{\rm gas}/\Sigma_{\rm stars}$ where stars are sub-dominant.

The resulting HI surface density $\Sigma_{\rm HI}(x,y)$ is then used to create face-on 
HI profiles $\Sigma_{\rm HI}(r)$
for all model galaxies. $R1$ is then determined as the radius, where $\Sigma_{\rm HI}(r)$ 
drops below $1\;M_{\odot}\rm pc^{-2}$
and the HI mass $M_{\rm HI}$ as the total mass out to the radius, 
where $\Sigma_{\rm HI}(r)$ drops below $0.2\;M_{\odot}\rm pc^{-2}$,
which corresponds to the detection threshold of the Bluedisk sample.
To consider the effect of beam smearing in observations, 
we convolve $\Sigma_{\rm HI}(x,y)$ with an elliptical Gaussian Kernel
with FWHM values of 14 and 9 kpc for major and minor axes. From the convolved 
surface density $\Sigma_{\rm HI, smooth}(x,y)$,
we then derive $\Sigma_{\rm HI, smooth}(r)$, $M_{\rm HI, smooth}$ and $R1_{\rm smooth}$.

\subsubsection{Results of the SPH simulation}

\begin{figure*}
\centering
\includegraphics[width=18cm]{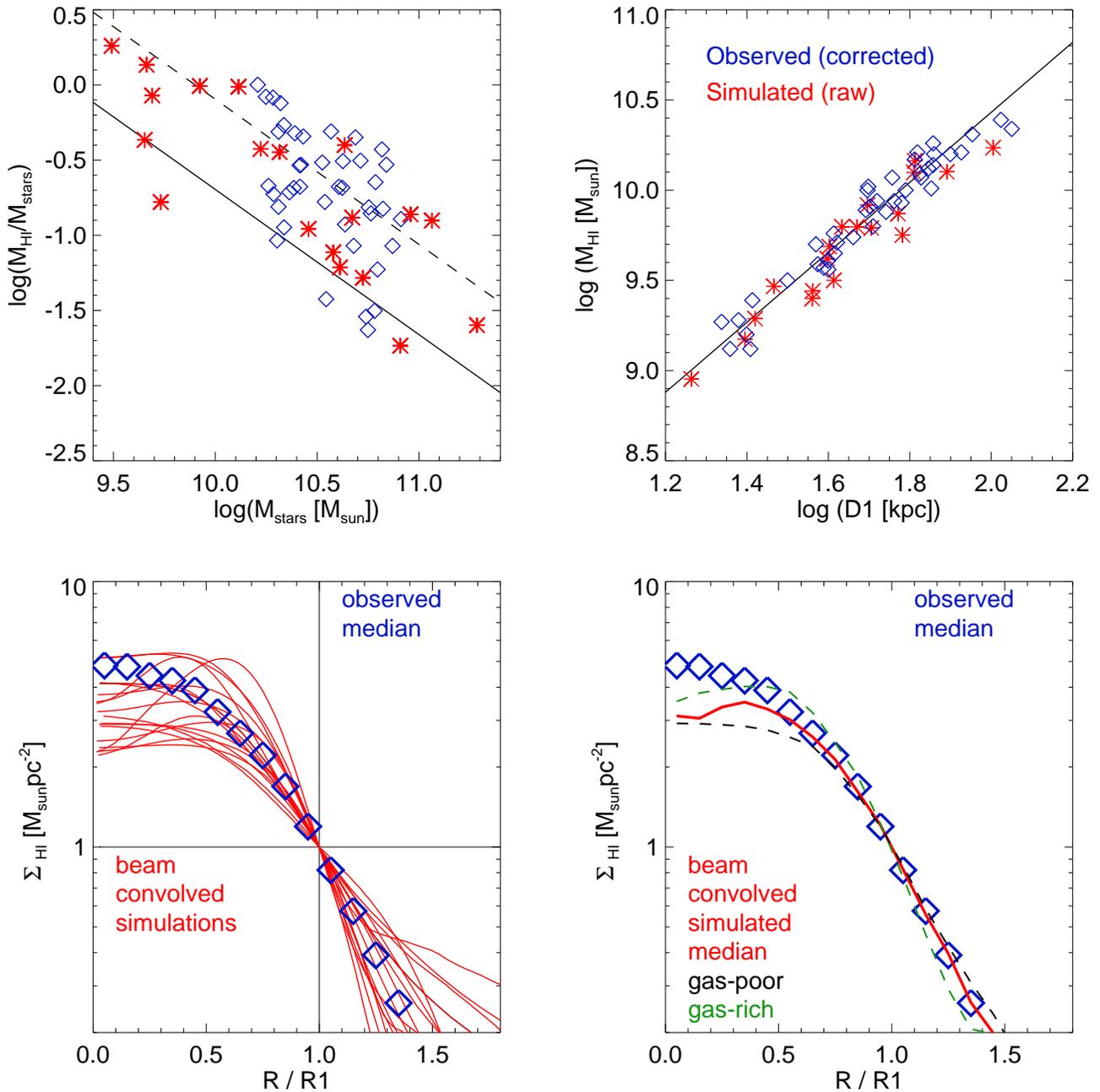}
\caption{
A comparison of the SPH simulations and the Bluedisk observations.
Panel 1: The HI-to-stellar mass ratio $M_{\rm HI}/M_{\rm stars}$ as a function of stellar mass $M_{\rm stars}$ for simulated
(red stars) and observed galaxies (blue diamonds). The solid black line shows the median
relation between HI mass fraction and stellar mass found by C10, the dashed black line
is offset from this relation by +0.6 dex.
Panel 2: The HI mass-size relation. For simulations (red stars) we plot $M_{\rm HI}$ and $D1 = 2\;{\rm R1}$ determined from the un-convolved
profiles, for observations (blue diamonds) we plot values that were corrected to account for beam effects.
Overplotted in black is the relation found by \citet{Broeils97}.
Panel 3: HI mass profiles for radii normalized to $R1$.
We plot the convolved profiles $\Sigma_{\rm HI, smooth}(r)$ for all 19 simulated disks (red lines).
We overplot the observed median profile (blue diamonds).
Panel 4: As Panel 3, but now we compare the simulated median profile (red line) to the 
observed median profile (blue diamonds).
We also show the medians for a gas-rich (green dashed) and a gas-poor (black dashed) sub-sample.}
\label{sim}
\end{figure*}

The results of a comparison between simulations and observations are depicted in Fig. \ref{sim}.

In the upper-left panel we depict the HI-to-stellar mass ratio $M_{\rm HI}/M_{\rm stars}$ as a function of stellar mass $M_{\rm stars}$
for the simulated and observed galaxies. The simulations (red) span a wider range in stellar mass ($9.5<\log(M_{\rm stars}/M_{\odot})<11.3$)
than the observations which were limited to $10.<\log(M_{\rm stars}/M_{\odot})<11$.
In terms of gas-richness, by construction about half of the observed galaxies lie above the black dashed line which
depicts the median relation found by C10 (solid black line) shifted up by 0.6 dex. Only 4 of 19 simulated galaxies are
above this line. The remaining  simulated galaxies have a similar distribution of HI-mass fraction as the       
control sample, which have lower gas fractions.

In the upper-right panel of Fig. \ref{sim} we depict the HI mass-size relation. 
For the observed galaxies, we use D1 values corrected for beam-smearing effects
and for simulated galaxies, we use   un-convolved
values. We see that the observed and the simulated  samples both follow the relation
found by \citet{Broeils97}. The agreement between the simulations and the
observations is very good . The simulations show a very mild
offset to lower HI masses, which may be connected to the difference in gas fraction of the samples or a minor problem
of the HI modeling. Note also that the range of sizes of the galaxies in the 
two samples is very similar. We find that the largest HI disks
in simulations form from gas that cools and settles in a disk after merger events at $z\sim 1$.

In the lower-left panel of Fig. \ref{sim} we depict a compilation of all 
19 beam-convolved simulated HI-mass profiles $\Sigma_{\rm HI, smooth}(R)$
with radius $R$ in units of $R1$. All profiles have  similar shape. They are rather flat in the centre and
fall exponentially at $R>0.75\; {\rm R1}$. The central values in surface density scatter between 
2 and 5 $M_{\odot}\rm pc^{-2}$, the
outer exponential scale-lengths are between 0.15 and 0.45 $R1$. We overplot with diamonds 
the observed median profile of the Bluedisk galaxies and note that
there is good agreement between simulations and observations.

The lower-right panel shows a comparison between the observed and the simulated 
median profiles. The central surface densities in simulations
are slightly lower than in the observations, consistent with the slightly lower 
HI gas fractions in the top left panel . The outer profile at $R>0.5\;{\rm R1}$ agrees
very well with the observations. We also note that the median outer HI scale-length 
of the un-convolved simulated disks is 0.17 in units of $R1$, which compares 
well to the beam-corrected mean value of 0.19 found for the Bluedisk sample.
The dashed green and black curves show the median profiles we obtain , when we use the offset of the HI-to-stellar mass ratio of the simulated galaxies from the solid black line in the upper-left panel of Fig. 10 to equally divide the simulated sample into a gas-rich (green) and a gas-poor (black) subsample.

The dashed green and black curves show the mean profiles for simulated galaxies divided by 
HI-to-stellar mass. Simulated galaxies falling above the dashed black line in the
top left panel are included in the ``gas-rich'' sub-sample, while those
falling below this line are included in the ``gas-poor'' sub-sample.

 We find that the gas-rich sample has higher central surface densities 
and shows steeper decline at the outer radii compared to the gas-poor sample. 
This differs from the observation that gas-richness has no influence on 
the outer slope. We caution however that we are looking at two samples 
with $\sim 10$ objects each. Considering the corresponding statistical
errors there is no significant disagreement between the outer slopes. 
Moreover, if we use means instead of medians , the difference in outer scale-length becomes very small.

\subsection{Comparison with semi-analytic  models}\label{sec:SAM}

In this section, we compare the observations with the semi-analytic models of galaxy 
formation of Fu et al. (2013; hereafter Fu13). 
In these models, each galaxy disk is divided into a series of concentric rings to trace 
the radial profiles of stars, interstellar gas and metals. 
Phyical prescriptions are adopted to partition the interstellar cold gas into atomic and molecular phases. 
In order to take into account the effects of beam smearing in the observations, 
the model radial profiles of $\Sigma_{\rm HI}$ are convolved by a Gaussian function with FWHM=10.6 kpc. 
The radius $R1$ is then derived from these smoothed profiles.
In order to carry out as fair a comparison with the observations as possible, we select the galaxy with $\log_{10}[M_{\rm HI}/\ms]>9.0$, $10.0<\log_{10}[M_{\rm HI}/\ms]<11.0$ and $-1.8<\log_{10}[M_*/M_{\rm HI}]<0.2$ from the model sample at $z=0$.

\subsubsection{Results from the Semi-analytical model comparison}

\begin{figure*}
\centering
 \includegraphics[angle=0,scale=0.72]{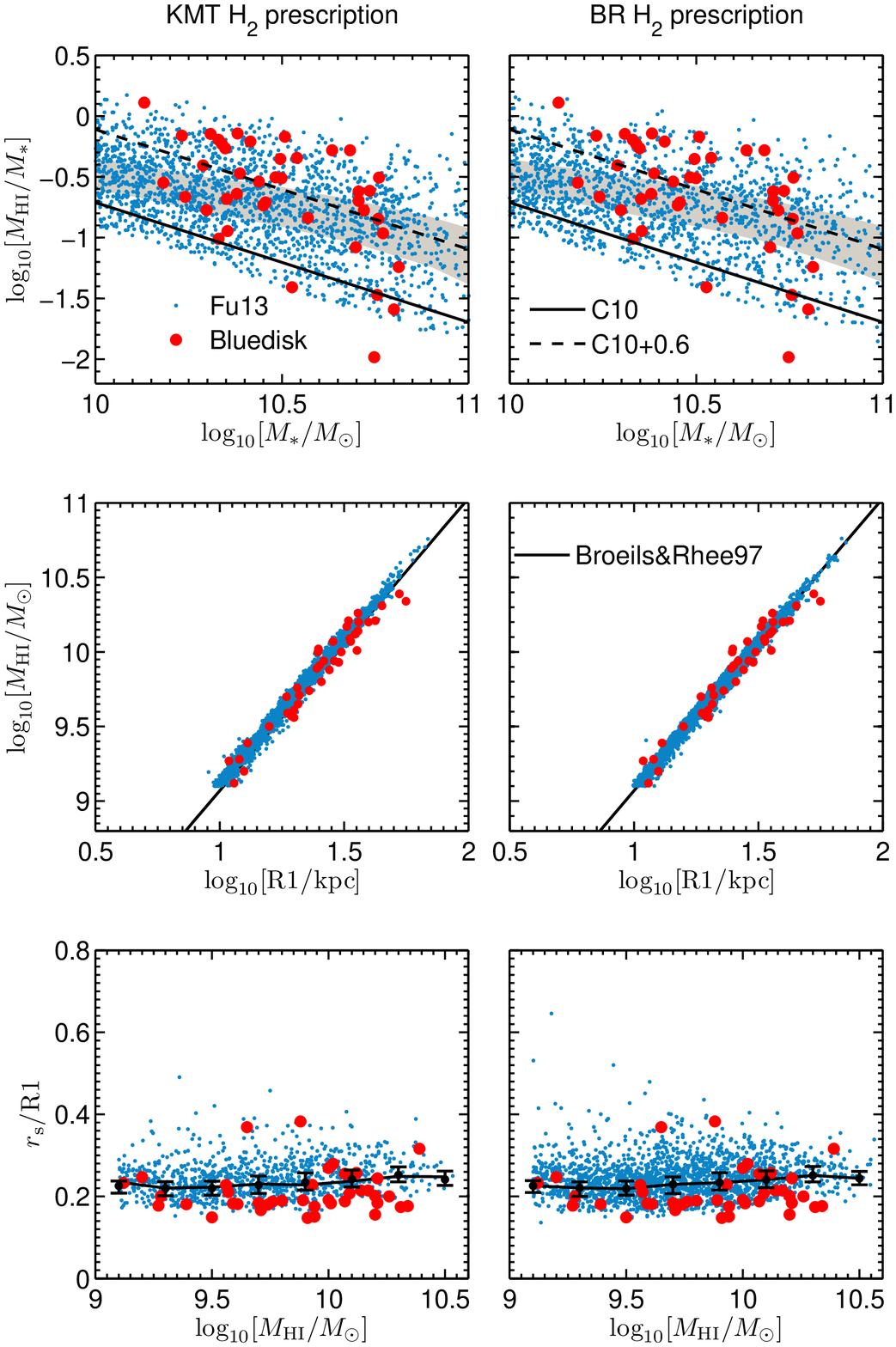}
 \caption
 {The comparison between the Fu13 model results and the Bluedisk observational sample. 
The two columns represent the Fu13 models with two H$_2$ fraction prescriptions
(see Fu13 for details). In each panel, the blue dots represent the model sample 
and the red dots are for the Bluedisk sample.
 Top two panels: The HI-to-stellar mass ratio as a function of stellar mass. 
The gray areas represent the $\pm1\sigma$ deviations around the mean values for the Fu13 model sample. For comparison, the black solid lines show the median relation in C10 and black dashed lines are offsets from solid lines by 0.6 dex.
 Middle two panels: The HI mass-size relation. The R1 values from observations 
are corrected for beam-smearing effects and R1 values from models are un-convolved. The black lines represent the fitting equation in Broeils \& Rhee (1997, the diameter $D1$ is converted to radius R1).
 Bottom two panels: The relation between $\rs/$R1 and HI mass. The black solid curves 
represent the model median values and the error bars represent the $\pm1\sigma$ deviations 
around the median values for the model sample. $\rs/$R1 values from models are un-convolved 
 and $\rs/$R1 values from Bluedisk observations are beam-corrected values.
 }\label{fig:fu13HI}
\end{figure*}

In the top two panels of Fig. \ref{fig:fu13HI}, we show the HI-to-stellar mass ratio as 
a function of stellar mass for the Fu13 model galaxies and the Bluedisk sample. 
The two panels represent the model results with two H$_2$ fraction prescriptions. 
The blue points correspond to individual model galaxies, while the red points 
correspond to individual Bluedisk galaxies. The grey shaded area shows the 1$\sigma$ 
deviation around the median relation between $M_{\rm HI}/M_*$  and $M_*$ for the 
model galaxies.
By construction, the model galaxies span the same range in stellar mass
and HI mass fraction as the Bluedisk sample.  
In the middle two panels of Fig. \ref{fig:fu13HI}, we show the HI mass-size relation
for the models and the data. The R1 values from the observations are corrected for 
beam-smearing and the model values are un-convolved. 
The model results fit both the Broeils \& Rhee (1997) and  the observational data extremely well.
We will discuss  the origin of the tight correlation between HI disk mass and 
size in the models in more detail in Section \ref{sec:mass-size}.
Finally, the bottom two panels show the
relation between  $\rs/$R1 and HI mass for both models and data.   
$\rs$ is defined as the scale-length of the HI profile at $r>0.75$R1 obtained by
fitting  an exponential profile to this region of the disk. The black solid curves show the 
model median values and the error bars represent the $\pm1\sigma$ deviations around the median. 
$\rs/$R1 is around 0.22 to 0.24 for model galaxies, compared to  0.19 for the Bluedisk sample.

\begin{figure}
\centering
 \includegraphics[angle=0,scale=0.48]{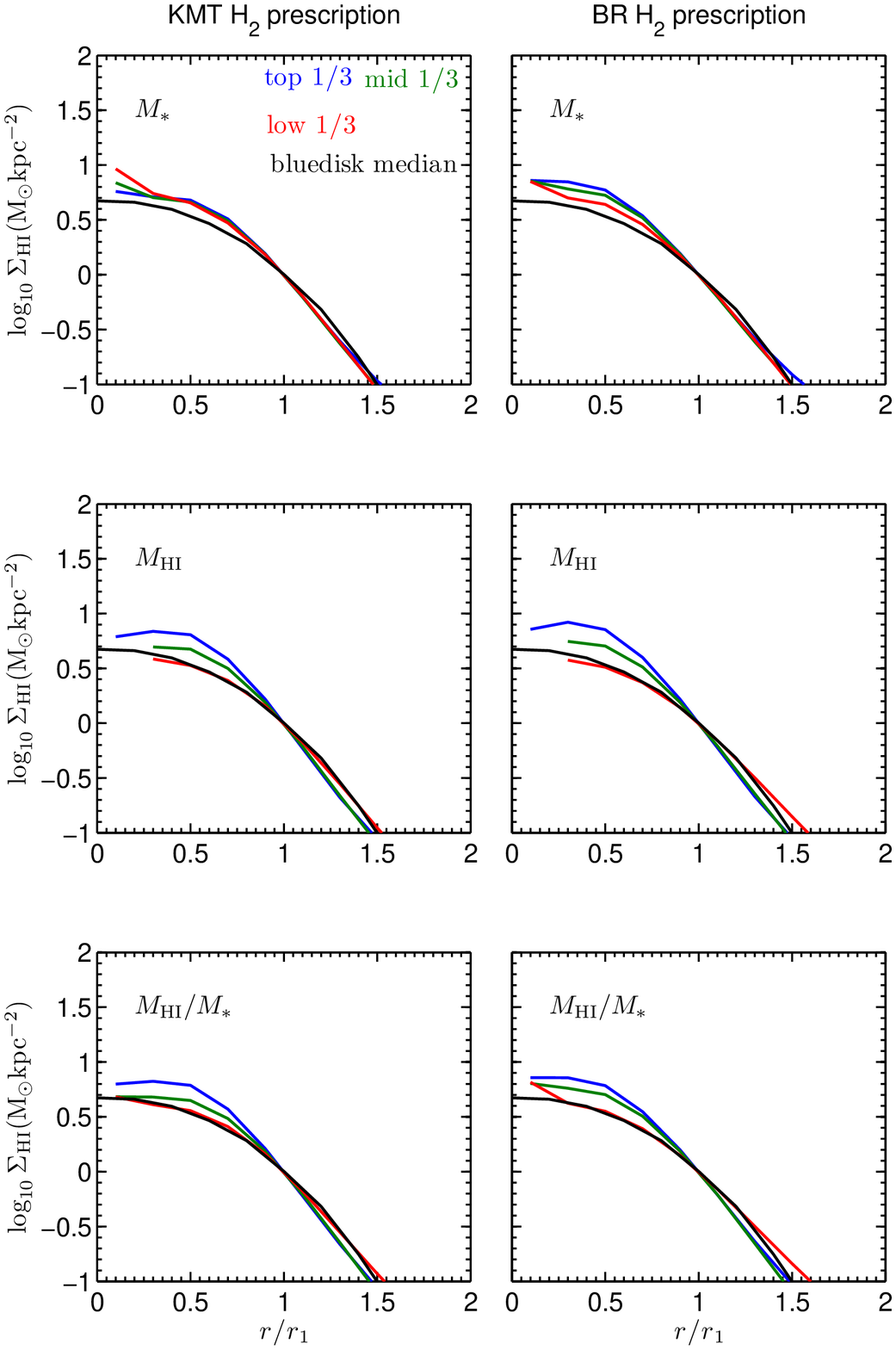}
 \caption
 {The relation between beam-convolved HI radial surface density profiles $\Sigma_{\rm HI}$ and the normalized radius $r/$R1 from the Fu13 model. The two columns represent the results of models with two H$_2$ fraction prescriptions. Similar to Fig. \ref{fig:avgprof_r1}, the model samples are equally divided into 3 parts by $M_*$, $M_{\rm HI}$ and $M_{\rm HI}/M_*$ from the top to bottom rows. In each panel, the red, green and blue curves represent the median profiles in the lowest to highest parameter bins. The black curves represent the median curves from the whole Bluedisk sample.
 }\label{fig:profiles}
\end{figure}

Finally, in Fig. \ref{fig:profiles}, we show the beam-convolved HI radial surface density profiles 
in units of $r/$R1 from the Fu13 models with two H$_2$ fraction prescriptions. 
Similar to Fig. \ref{fig:avgprof_r1}, we divide model samples into 3 equal parts
according to  stellar mass $M_*$, HI mass $M_{\rm HI}$, and HI-to-stellar mass ratio $M_{\rm HI}/M_*$, 
from the top to bottom rows. In each panel, the red, green and blue curves represent the 
median model profiles for the highest,intermediate and lowest
values of each parameter.   The black curves show the median 
profile of the  Bluedisk sample. 

As can be seen, the model and the observational profiles agree extremely well in the outer disk.
The HI density profiles in the inner disk are systematically too high compared to observations.
When the sample is split according to HI mass or HI mass fraction, the same {\em qualitative}
trends are seen in both models and data. We will explain the origin of these trends in the
next section.

\section{Interpretation of the results} \label{sec:explain}

\subsection{Outer HI profiles in semi-analytic models} \label{sec:outsams}

We have demonstrated that the outer disk HI surface density profiles have universal
form when scaled to a radius corresponding to a fixed HI surface density and that the  
slope of the outer disk  $\rs/$R1 is almost constant among different galaxies. 

In the Fu13 models, 
newly infalling gas is assumed to be distributed exponentially and is directly superposed 
onto the pre-existing gas profile from the previous time step.
Gas also flows towards the centre of the disk with inflow
velocity $v_{\rm inflow} = \alpha r$ ($v_{\rm inflow} \sim$  7 km/s 
at a galactocentric radius of 10 kpc) .  Following the prescription
in Mo, Mao \& White (1998), the scale length of the exponential infalling gas 
is given by $r_{\rm cool}=\left(\lambda/\sqrt 2 \right)r_{\rm vir}$,
in which $\lambda$ and $r_{\rm vir}$ are the spin parameter and virial radius of
the  galaxy's dark matter halo. Since the increase of $r_{\rm vir}$ and $r_{\rm cool}$
scales with the age of the Universe, the outer disk gas profiles are mainly determined 
by recent gas accretion. This disk galaxy  growth paradigm is called 
``inside-out'' disk formation (e.g Kauffmann et al. 1996; Dutton 2009; Fu et al. 2009), 
and an illustration of inside-out growth of the gas disk can be 
found in Fig. 1 in Fu et al. (2010).

According to the definition of outer disk scale-length $\rs$, the outer disk 
HI surface density profile can be written as
\begin{equation}
\Sigma_{\rm{HI}}=\Sigma_0\exp \left( -\frac{r}{\rs} \right)
\label{eq:sgas}
\end{equation}
in which $\Sigma_0$ is disk central HI surface density extrapolated from the outer disk HI profile. 
Let us suppose that  $\rs$ in Eq. (\ref{eq:sgas}) is  approximately equal to
 $r_{\rm infall}$ at the present day, and $\Sigma_0$ represents the 
amount of gas accreted at $r=0$ in the recent epoch.

Substituting $1\mspc=\Sigma_0\exp(-{\rm R1}/\rs)$ into Eq. (\ref{eq:sgas}),
$\rs/$R1 can be written as
\begin{equation}
\frac{{{r_{\rm{s}}}}}{{{{\rm R1}}}} = \frac{1}{{\ln \left[ \Sigma_0/{M_ \odot }{\rm{pc}}^{ - 2} \right]}}.
\label{eq:rsr1}
\end{equation}
Eq. (\ref{eq:rsr1}) implies that $\rs/$R1 actually relates to the amount of 
recent gas accretion at $r=0$. For simplicity, 
we define $\dot\Sigma_0=d\Sigma_0/dt$ as the cold gas infalling rate at $r=0$.

\begin{figure*}
\centering
 \includegraphics[angle=-90,scale=0.65]{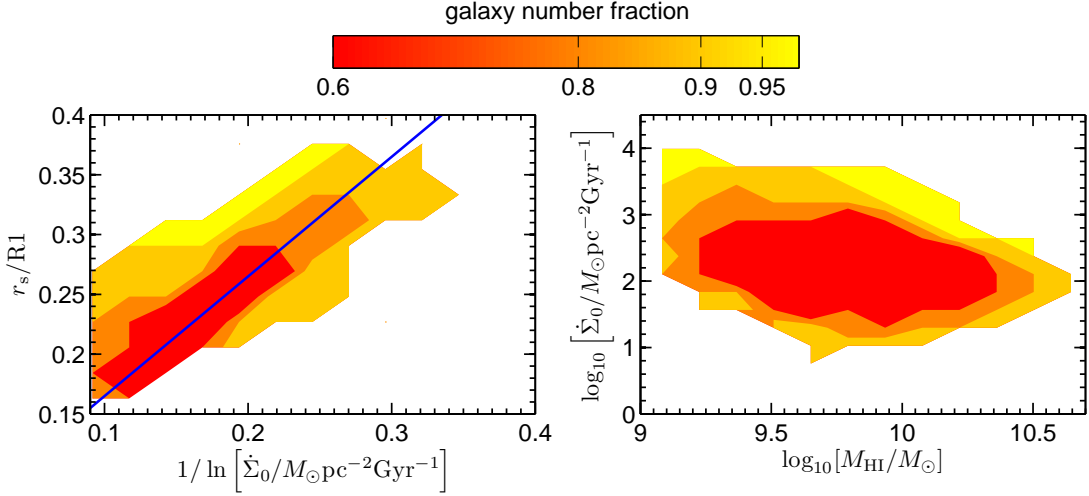}
 \caption
 {Left panel: the relation between $\rs/$R1 and $1/\ln\left[\dot\Sigma_0/M_\odot\rm{pc^{-2}Gyr^{-1}} \right]$ 
for model galaxies. $\dot\Sigma_0$ is the gas infall rate in the disk centre averaged
over the past 0.5 Gyr and $\rs$ is the exponential scale length of the outer disk. 
The blue solid line is the fitting line Eq. (\ref{eq:fitting}). 
Right panel: the relation between $\dot\Sigma_0$ and $M_{\rm HI}$ for model galaxies. 
In both panels, the galaxies are from model results with KMT H$_2$ fraction 
prescription at $z=0$. The contours indicate the fraction of model galaxies 
located in a given region of parameter space, as given by the colour key at the top of the plot.
 }\label{fig:sigma0}
\end{figure*}

We have tested the correlation between $\rs/$R1 and $\dot\Sigma_0$ averaged
over different timescales.  We find the 
best correlation with  $\rs/R1$ if  $\dot\Sigma_0$ is averaged over a timescale
corresponding to the past 0.5 Gyr. In the left panel of Fig. \ref{fig:sigma0}, 
we plot the relation between $\rs/R1$ and $1/\ln\dot\Sigma_0$ (averaged in the recent 0.5 Gyr) 
for model galaxies with $M_*>10^{10}\ms$ and $M_{\rm HI}>10^9\ms$, based on KMT H$_2$ prescription. The blue solid line 
represents the fitting equation
\begin{equation}
\rs/{\rm R1} = 0.065+\left(\ln\left[\dot\Sigma_0/M_\odot\rm{pc^{-2}Gyr^{-1}} \right]\right)^{-1}.
\label{eq:fitting}
\end{equation}
However, we note that a correlation between $\rs/$R1 and HI excess ($\dfHI$) is not found from the observation (section~\ref{sec:uniprof}).

The right panel of Fig. \ref{fig:sigma0} shows the relation between $M_{\rm HI}$ 
and $\dot\Sigma_0$. There is no relation between the two quanties, i.e. HI mass alone
cannot be used to infer the recent accretion rate of a galaxy. 

To summarize: the universal outer disk profiles in the semi-analytic models
originate from the combination of the {\em assumption} that infalling gas
has an exponential profile and of the inside-out growth of disks. 

\subsection{Inner HI profiles in semi-analytic models} \label{sec:mass-size}

\begin{figure*}
\centering
  \includegraphics[angle=-90,scale=0.65]{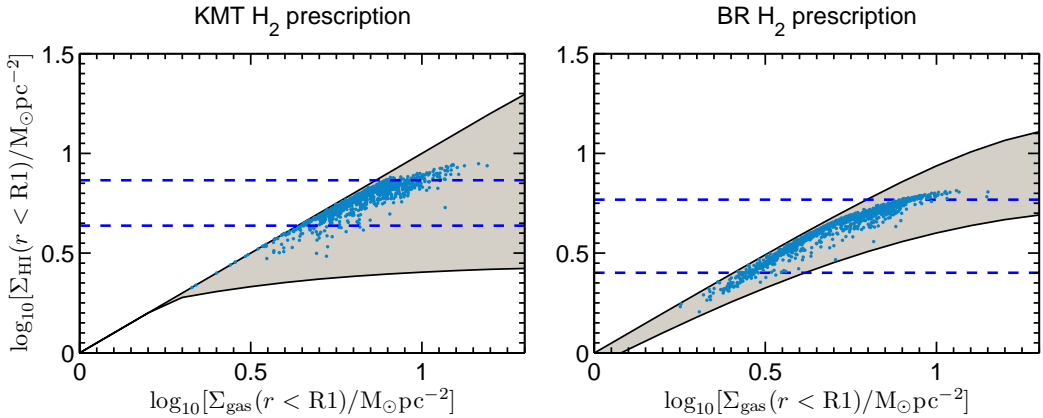}\\
  \caption{The relation between inner disk ($r<$R1) HI surface density and total cold gas surface density from Fu13 models.
  The blue dots in two panels are model samples with two $\h2$ prescriptions. The two dashed lines in each panel indicate the values of $\Sigma_{\rm HI}(r<{\rm R1})$ that contain $90 \%$ percent of the model samples. The gray areas represent the relations of $\Sigma_{\rm HI}$ vs. $\Sigma_{\rm gas}$ results directly calculated from the two $\h2$-HI transition prescriptions in Fu13.}
  \label{fig:sgassh1}
\end{figure*}

In  the inner part ($r<$R1) of the disks, the HI surface density is mainly determined 
by the conversion between atomic and molecular gas. Higher total gas surface density 
($\sgas=\SHI+\Sh2$) leads to higher $\h2$-to-HI ratio, which maintains  the value of $\SHI$ in a narrow range. 
In Fig. \ref{fig:sgassh1} , we plot HI surface density as a function of   total cold gas 
surface density in the inner disks for model galaxies. We define 
$\SHI(r<{\rm R1})=M_{\rm HI}(r<{\rm R1})/\pi {\rm R}{1^2}$ and $\sgas(r<{\rm R1})=M_{\rm HI+\h2}(r<{\rm R1})/\pi {\rm R}{1^2}$. 
We show results for two $\h2$ fraction prescriptions. In each panel, the dashed lines indicate 
the values of $\Sigma_{\rm HI}(r<{\rm R1})$ that enclose  90\% percent of the model galaxies . 
The gray areas indicate the region of the  $\Sigma_{\rm gas}$ versus  $\Sigma_{\rm HI}$ plane that  
is plausibly spanned by the two atomic-to-molecular gas transition prescriptions. For the KMT 
$\h2$ prescription, we adopt a range in gas metallicity relative to the 
solar value $\rm{[Z/H]_{gas}}$ from -0.5 to 1. For the BR $\h2$ prescription, we adopt
a range in stellar surface density $\sqrt{\Sigma_*\Sigma^0_*}$ 
(see Equation 32 in Fu et al. 2010) from $1~\mspc$ to $4000~\mspc$.
We note that in practice  90 \% of the model galaxies have $\Sigma_{\rm HI}(r<{\rm R1})$ 
values between $ \sim 3$ and $ \sim 4\mspc$. 

To summarize: it is the combination of a narrow range in inner HI surface density and 
a ``Universal'' outer exponential profile  that leads to the very tight HI mass-size relation.
We note that 
the atomic-to-molecular gas conversion plays the more important role, 
because the inner disk contains a larger fraction of the total HI gas compared to the outer disk.

\subsection{Main caveats} We note that the assumption that gas accreted from the halo is distributed
on the disk with an exponential profile has no {\em a priori} 
physical justification. Bullock et al (2001)
studied the angular momentum profiles of a statistical sample of halos drawn from
high-resolution N-body simulations of the $\Lambda$CDM cosmology and showed that 
the cumulative mass distribution of specific angular momentum $j$ in  halos could be
fit with a function that follows a power law  and then flattens at large $j$. 

They explored implications of their M($< j$) profile on the formation of galactic disks 
assuming that $j$ is conserved during infall. They showed
that the implied gas density profile deviates from an exponential disk, 
with higher densities at small radii and a tail extending to large radii. 

It is thus something of mystery why the hydrodynamical simulations presented in this
paper have outer disks that agree so well with the observed ``universal'' outer exponential
profiles seen the data. The solution likely lies in the complicated interplay between
the infall of new gas, star formation, supernova feedback and gas inflow processes 
occurring in the simulation and will
form the subject of future work.

\section{Summary}

We have measured the azimuthally-averaged radial profiles of HI gas in 42 galaxies 
from the Bluedisk sample. We investigated how the shape of HI profiles vary as a function 
of galactic properties.  We developed a model to describe the shape of HI profiles 
which is an exponential function of radius in the outer regions and has a depression 
towards the center. We derive  maximum-likelihood estimates of $\rs$, the
scale radius  of the outer exponential HI disk. By inverting the relation between
star formation rate and $\h2$ surface density, we also derive estimates of the
total gas surface density profiles of our galaxies. Finally, we compare our 
observational results with predictions from SPH simulations and semi-analytic
models of galaxy formation in a $\Lambda$CDM universe. The main results can
be summarized as follows: 

\begin{enumerate}
\item The HI disks of galaxies exhibit a homogeneous radial distribution of HI  in their outer regions, 
when the radius is scaled to R1, 
the radius where the column density of the HI is 1$\ms$ pc$^{-2}$. 
The outer distribution of HI is well-fit by an exponential function with a scale-length 
of 0.18 R1. This function does not depend on stellar mass, stellar mass surface density, 
optical concentration index, global NUV-r colour, color gradient, HI mass, 
HI-to-stellar mass fraction, HI excess parameter, or HI disk morphologies.

\item By comparing the radial profiles of the HI-rich
galaxies with those of the control systems, we deduce that 
in about half the galaxies, most of the excess gas lies outside the stellar disk, in the exponentially
declining outer regions of the HI disk. In the other half, the excess
is more centrally peaked.  

\item  The median total cold gas profile of the galaxies in our sample
agrees well with the ``universal'' radial profile proposed  by Bigiel \& Blitz (2012).  
However, there is considerable scatter around the B12 parametrization, particularly in the
outer regions of the disk.

\item Both the SPH simulations and Semi-analytical models are able 
to reproduce the homogeneous profile of HI in the outer region with the correct scale-length. 

\item
In the semi-analytic models, the universal shape of the outer HI radial profiles is a
consequence of the {\em assumption} that infalling gas is always distributed
exponentially. The conversion of atomic gas to molecular form explains
the limited range of HI surface densities in the inner disk. These two factors
produce the tight HI mass-size relation. 

\end{enumerate}

\section*{Acknowledgements}

GALEX (Galaxy Evolution Explorer) is a NASA Small Explorer, launched in April 2003, developed in cooperation with the Centre National d'\'{E}tudes Spatiales of France and the Korean Ministry of Science and Technology.

Funding for the SDSS and SDSS-II has been provided by the Alfred P. Sloan Foundation, the Participating Institutions, the National Science Foundation, the U.S. Department of Energy, the National Aeronautics and Space Administration, the Japanese Monbukagakusho, the Max Planck Society, and the Higher Education Funding Council for England. The SDSS Web Site is http://www.sdss.org/.

Jian Fu acknowledges the support from the National Science Foundation of China No. 11173044 and the Shanghai Committee of Science and Technology grant No. 12ZR1452700.

MA acknowledges support from the DFG Excellence Cluster ``Origin and Structure of the UniverseÓ.
\bibliographystyle{mn2e}

\end{document}